\documentstyle[11pt,subeqn]{article}
\begin{document}
\begin{titlepage}

\begin{center}
{\huge\bf Gravitational Waves in the
Nonsymmetric Gravitational Theory\\}
\vspace{1in}
{\bf N. J. Cornish, J. W. Moffat and D. C. Tatarski\\}
{\bf Department of Physics\\}
{\bf University of Toronto\\}
{\bf Toronto, Ontario M5S 1A7, Canada\\}
\vspace{1in}
{\bf\large Abstract}
\end{center}

We prove that the flux of gravitational radiation from an isolated source
in the Nonsymmetric Gravitational Theory is identical to that found in
Einstein's General Theory of Relativity.

\vspace{1in}

\centerline{{\bf UTPT-92-18}}

\vspace{.2in}

\centerline{{\bf GR-QC/9211023}}

\vspace{.4in}

\centerline{{\bf November 1992}}

\end{titlepage}

\section{Introduction}

It has recently been claimed by Damour, Deser and McCarthy
\cite{DDM} that the Nonsymmetric Gravitational Theory (NGT) \cite{Moff91}
is unphysical due to radiative instability. They claim that NGT predicts
an infinite flux of negative energy from gravitational waves. We prove this
claim to be false by demonstrating that the flux of gravitational
radiation from an isolated source in NGT is identical to that found in General
Relativity (GR).

Our proof is based on the following analysis. Firstly, we solve the NGT field
equations for an isolated axi-symmetric source using the expansion technique
employed in \cite{DDM}. (The exact version of this work will be published
elsewhere \cite{CMT}). We find that the skew sector of the theory does not
contribute to the flux of gravitational radiation due to the short-range
nature of the skew fields. Secondly, since the NGT wave equations allow
linear superposition in the wave zone, we see that any compact source has
a wave field that can be expressed as the superposition of a suitable
collection of differently aligned axi-symmetric solutions. This completes
our proof.

\section{NGT Vacuum Field Equations} \label{fielde}
The NGT Lagrangian without sources takes the form:
\begin{equation} \label{lagr}
{\cal L} =  \sqrt{-g}g^{\mu \nu} R_{\mu \nu}(W),
\end{equation}
with \(g\) the determinant of \(g_{\mu \nu}\). The NGT Ricci tensor
is defined as:
\begin{equation} \label{ricciw}
R_{\mu \nu}(W) = W^{\beta}_{\mu \nu , \beta}- \frac{1}{2}
(W^{\beta}_{\mu \beta , \nu}+W^{\beta}_{\nu \beta , \mu}) -
W^{\beta}_{\alpha \nu}W^{\alpha}_{\mu \beta}+W^{\beta}_{\alpha
\beta}W^{\alpha}_{\mu \nu},
\end{equation}
where \(W^{\lambda}_{\mu \nu}\) is an unconstrained nonsymmetric
connection :
\begin{equation} \label{connw}
W^{\lambda}_{\mu \nu}=W^{\lambda}_{(\mu \nu)}+W^{\lambda}_{[\mu
\nu]}.
\end{equation}
(Throughout this paper parentheses and square brackets enclosing
indices stand for symmetrization and antisymmetrization,
respectively.)
The contravariant nonsymmetric tensor \(g^{\mu \nu}\) is defined in
terms of the equation:
\begin{equation} \label{inverse}
g^{\mu \nu} g_{\sigma \nu}=g^{\nu \mu} g_{\nu
\sigma}=\delta^{\mu}_{\sigma}.
\end{equation}

If we define the torsion vector as:
\begin{equation}
W_{\mu} \equiv W^{\nu}_{[\mu \nu]} = \frac{1}{2} \left(
W^{\nu}_{\mu \nu}- W^{\nu}_{\nu \mu} \right),
\end{equation}
then the connection $\Gamma^{\lambda}_{\mu \nu}$, where
\begin{equation} \label{conng}
\Gamma^{\lambda}_{\mu \nu} = W^{\lambda}_{\mu \nu} + \frac{2}{3}
\delta^{\lambda}_{\mu} W_{\nu} \; ,
\end{equation}
is torsion free:
\begin{equation} \label{gtors}
\Gamma_{\mu} \equiv \Gamma^{\alpha}_{[\mu \alpha]} = 0.
\end{equation}
Defining now:
\begin{equation} \label{riccig}
R_{\mu \nu}(\Gamma) = \Gamma^{\beta}_{\mu\nu,\beta} -
\frac{1}{2}(\Gamma^{\beta}_{(\mu\beta),\nu} + \Gamma^{\beta}_{(\nu
\beta) , \mu})-\Gamma^{\beta}_{\alpha\nu}\Gamma^{\alpha}_{\mu
\beta}+\Gamma^{\beta}_{(\alpha\beta)}\Gamma^{\alpha}_{\mu\nu},
\end{equation}
we can write:
\begin{equation} \label{ricciw=g}
R_{\mu \nu}(W) = R_{\mu \nu}(\Gamma) + \frac{2}{3} W_{[\mu ,\nu]},
\end{equation}
where \(W_{[\mu,\nu]}=\frac{1}{2}(W_{\mu,\nu}-W_{\nu,\mu})\).
Finally, the NGT vacuum field equations can be expressed as:
\begin{subequations} \label{fens}
\begin{equation} \label{fensgamma}
g_{\mu\nu,\sigma} - g_{\rho\nu} {\Gamma}^{\rho}_{\mu\sigma} -
g_{\mu\rho} {\Gamma}^{\rho}_{\sigma\nu} = 0 ,
\end{equation}
\begin{equation} \label{fensdiver}
{(\sqrt{-g}g^{[\mu \nu]})}_{ , \nu} = 0 ,
\end{equation}
\begin{equation} \label{fensricci}
R_{\mu \nu}(\Gamma) = \frac{2}{3} W_{[\nu , \mu]}.
\end{equation}
\end{subequations}
It is convenient to decompose \(R_{\mu\nu}\) into standard symmetric and
antisymmetric parts: \(R_{(\mu\nu)}\), \(R_{[\mu\nu]}\), and then
rewrite the field equation (\ref{fensricci}) in the following form:
\begin{subequations} \label{fensriccis}
\begin{equation} \label{sym}
R_{(\mu \nu)}(\Gamma) = 0,
\end{equation}
\begin{equation} \label{asym}
R_{\{[\mu \nu] , \rho\} }(\Gamma)= R_{[\mu\nu],\rho}(\Gamma)
 + R_{[\nu\rho],\mu}(\Gamma)+ R_{[\rho\mu],\nu}(\Gamma) = 0 \; .
\end{equation}
\end{subequations}

We shall base our analysis on the expansion used in \cite{DDM}, where the
field equations are expanded in powers of the antisymmetric part,
$h_{[\mu\nu]}$, of the metric but where the symmetric part is taken to
be an exact GR background. The field equations become to lowest order:
\begin{subequations} \label{hexpan}
\begin{eqnarray}
 ^{{GR}}R_{\mu\nu}&=&0 \; , \\
D^{\alpha}F_{\mu\nu\alpha}+4\; ^{GR}R^{\alpha\;\;\beta}_{\;\; \mu\;\; \nu}
h_{[\alpha\beta]}&=&{4 \over 3}W_{[\nu,\mu]} \; , \label{srex} \\
D^{\mu}h_{[\mu\nu]}&=&0 \; ,
\end{eqnarray}
\end{subequations}
where $F_{\mu\nu\alpha}$ is the cyclic curl of $h_{[\mu\nu]}$ and $D^{\mu}$
is the GR background covariant derivative.

For our current purposes, we shall be using the radiative, axi-symmetric
GR background found by Bondi, van der Burg and Metzner \cite{BBM}. Since
their result is given as an expansion in inverse powers of the radial
coordinate, we shall also have to expand the above equations in the same
way. The two expansions are perfectly compatible, and both expansions are well
suited to studying the equations in the wave zone.

\section{The Metric} \label{coord}

The physical situation that we shall be studying is that of
an isolated spherical body which has been deformed by a perturbing
compression along a single axis. We seek to study the flux of gravitational
radiation emitted as the body oscillates through the cycle
oblate--spherical--prolate--spherical--oblate.

Due to the physical picture sketched above, and to the fact that we
are interested in the asymptotic behaviour of the field at spatial
infinity (in an arbitrary direction from our isolated source), polar
coordinates \(x^{0}=u, {\bf x} = (r,\theta,\phi)\) are the natural
choice. The ``retarded time'' \(u=r-t\) has the property that the
hypersurfaces with \(u=\mbox{constant}\) are light--like. Detailed
discussion of the coordinate systems permissible for investigation
of outgoing gravitational waves from isolated systems can be found
in \cite{BBM,Sachs}.

The covariant GR metric tensor corresponding to the situation
described above:
\begin{equation} \label{grmetric}
^{GR}g_{\mu\nu} = \left( \begin{array}{cccc}
Vr^{-1}e^{2\beta}-U^{2}r^{2}e^{2\gamma} & e^{2\beta} &
Ur^{2}e^{2\gamma} & 0 \\
e^{2\beta} & 0 & 0 & 0 \\
Ur^{2}e^{2\gamma} & 0 & -r^{2}e^{2\gamma} & 0 \\
0 & 0 & 0 & -r^{2}e^{-2\gamma}\sin^{2}\theta \end{array} \right),
\end{equation}
with \(U,V,\beta,\gamma\) being functions of \(u,r\) and
\(\theta\) was first given in \cite{Bondi60}. We shall be using the
forms found for the GR metric functions, and all derived quantities, as
given in \cite{BBM}.

In order to find the NGT generalization of the metric tensor
(\ref{grmetric}), we require that the symmetric part of the NGT
metric tensor is formally the same as the GR metric tensor. We
then impose the spacetime symmetries of the symmetric metric on
the antisymmetric tensor $g_{[\mu\nu]}$. This is achieved by demanding that
$\pounds_{ {\vec{\xi}_{(i)}} }g_{[\mu\nu]}=0$, where the Killing vector
field $\vec{\xi}_{(i)}$ is obtained from $\pounds_{ {\vec{\xi}_{(i)}} }
g_{(\mu\nu)}=0$. The solution to this equation for the metric (\ref{grmetric})
yields the single Killing vector field, $\vec{\xi}_{(1)}=\xi^{3}_{(1)}
\partial_{\phi}=\sin^{2}\theta \partial_{\phi}$. Imposing
$\pounds_{{\vec{\xi}_{(1)}} }g_{[\mu\nu]}=0$ yields:
\begin{equation}
\xi^{3}_{(1)}\partial_{\phi} g_{[\mu\nu]} +g_{[\mu 3]}\partial_{\nu}
\xi^{3}_{(1)} +g_{[3\nu]}\partial_{\mu}\xi^{3}_{(1)}=0 \; .
\end{equation}
This equation gives $\partial_{\phi}g_{[\mu\nu]}=0$, but does not exclude
any components of $g_{[\mu\nu]}$. This is markedly different from
the static spherically symmetric case, where the above procedure excludes
four of the six components of $g_{[\mu\nu]}$.

To make the problem tractable, we need to determine which $g_{[\mu\nu]}$
should be excluded. The criteria to be applied are that our solution
must reduce to the usual NGT static spherically symmetric solution when
passing through the equilibrium position, and that for each $g_{[\mu\nu]}$
set to zero, a corresponding field equation must also vanish.
To accomplish this, we note that the imposition of axi-symmetry splits
the antisymmetric field equations (\ref{fensdiver}), (\ref{asym})
into two sets of three independent equations. (This can be seen directly
from the block-diagonal form of the GR metric). The first set explicitly
involves the three functions $g^{[01]}\, , g^{[02]}\, , g^{[12]}$:
\begin{eqnarray}
\left(\sqrt{-g}g^{[\mu\nu]}\right)_{,\nu}&=&0 \hspace{1in}
(\mu=0,1,2) \; ,  \nonumber \\ \nonumber \\
R_{[01,2]}&=&0 \; .
\end{eqnarray}
These four equations are not independent owing to the existence of
the one identity:
\begin{equation}
\left(\sqrt{-g}g^{[\mu\nu]}\right)_{,\nu,\mu}=0 \hspace{1in} (\mu,\nu=0,1,2)
\; . \label{ident}
\end{equation}
The second set of four equations explicitly involves the three
functions $g^{[30]}\, ,g^{[31]}\, ,g^{[32]}$:
\begin{eqnarray}
\left(\sqrt{-g}g^{[3 \nu]}\right)_{,\nu}&=&0 \; , \nonumber \\
\nonumber \\
R_{[3 \mu ,\nu]}&=&0 \; .
\end{eqnarray}
These four equations are also not independent due to the one
identity:
\begin{equation}
\epsilon^{3 \mu\nu\rho}R_{[3 \mu ,\nu],\rho}=0 \; .
\end{equation}
We note that eliminating one complete set of three functions simultaneously
eliminates the three corresponding equations. However, eliminating one or
two functions from either set fails to reduce the number of equations and
leads to an over constrained system. Moreover, a Killing vector
analysis performed at the instant when the system passes through its
equilibrium position, shows that the set $\{ g^{[01]}\, , g^{[02]}\, ,
g^{[12]} \}$ must reduce to $g^{[01]}$, while the set $\{g^{[30]}\,
,g^{[31]}\, ,g^{[32]} \}$ must reduce to $g^{[32]}$. Since we know from
the exact spherically symmetric solution that asymptotic flatness requires
$g^{[32]}=0$, we must eliminate the entire second set of three functions
$\{g^{[30]}\, ,g^{[31]}\, ,g^{[32]} \}$.

In view of the above, the NGT generalization of the metric tensor
(\ref{grmetric}) is:
\begin{equation} \label{genmetric}
g_{\mu\nu} = \left( \begin{array}{cccc}
Vr^{-1}e^{2\beta}-U^{2}r^{2}e^{2\gamma} & e^{2\beta} + \omega &
Ur^{2}e^{2\gamma} + \lambda & 0 \\
e^{2\beta} - \omega & 0 & \sigma & 0 \\
Ur^{2}e^{2\gamma} - \lambda & -\sigma & -r^{2}e^{2\gamma} & 0 \\
0 & 0 & 0 & -r^{2}e^{-2\gamma}\sin^{2}\theta \end{array} \right),
\end{equation}
where \(\omega ,\lambda\) and $\sigma$ are functions of \(u,r\) and
\(\theta\). The contravariant metric tensor is given to order $h^{2}$ (where
$h=\omega,\lambda$ or $\sigma$) by:
\begin{equation} \label{contrmetric}
g^{\mu\nu} = \left( \begin{array}{cccc}
0 & e^{-2\beta}+g^{[01]}
& g^{[02]}  & 0 \\
e^{-2\beta}+g^{[10]} & e^{-2\beta}Vr^{-1}
& e^{-2\beta}U+ g^{[12]} & 0 \\
g^{[20]} &
e^{-2\beta} U +g^{[21]} & -e^{-2\gamma} r^{-2} & 0 \\
0 & 0 & 0 & -r^{-2}e^{2\gamma}\sin^{-2}\theta \end{array} \right),
\end{equation}
where
\begin{eqnarray}
g^{[01]}&=&-(\omega-\sigma U)e^{-4\beta} , \nonumber \\
g^{[02]}&=&-\sigma e^{-2\beta-2\gamma}r^{-2} ,\nonumber \\
g^{[12]}&=&-\left(Ur^{2}(\sigma U-\omega)e^{-4\beta}+e^{-2\beta-2\gamma}
(\lambda-\sigma V r^{-1})\right)r^{-2} , \nonumber\\
g&=&{\rm det}\, (g_{\mu\nu})=-r^{4}\sin^{2}\theta\, e^{4\beta} . \nonumber
\end{eqnarray}

\section{Solving the Field Equations}

We shall solve the field equations (\ref{hexpan}) using the GR background
given by \cite{BBM}:
\begin{subequations} \label{expansions}
\begin{equation} \label{expbeta}
\beta=-\frac{1}{4}c^{2}r^{-2}+...
\end{equation}
\begin{equation} \label{expgamma}
\gamma=cr^{-1}+\left(C-\frac{1}{6}c^{3}\right)r^{-3}+... ,
\end{equation}
\begin{equation} \label{expU}
U=-(c_{,\theta}+2c\cot\theta)r^{-2} + (2N + 3cc_{,\theta} +
4c^{2}\cot\theta) r^{-3} + ... ,
\end{equation}
\begin{equation} \label{expV}
V=r-2M - \left[N_{,\theta} +N\cot\theta -{c_{,\theta}}^{2} -
4cc_{,\theta}\cot\theta -\frac{1}{2}c^{2}(1+8\cot^{2}\theta)
\right] r^{-1} + ...  ,
\end{equation}
\end{subequations}
where \(c(u,\theta),N(u,\theta),M(u,\theta)\) are functions of
integration and \(C(u,\theta)\) satisfies:
\[
4C_{,u}=2c^{2}c_{,u}+2cM+N\cot\theta-N_{,\theta}.
\]
We begin by solving the three field equations $
\left(\sqrt{-g}g^{[\mu\nu]}\right)_{,\nu}=0, \hspace{.1in}
(\mu=0,1,2) $ with the skew functions expanded in inverse powers of
$r$. To this end, it is convenient to work with the following linear
combinations of $\omega, \lambda$ and $\sigma$:
\begin{subequations}
\begin{eqnarray}
\eta &=& (\omega-\sigma U)e^{-2\beta} \; , \\
\kappa &=&  (\lambda -\sigma V r^{-1})e^{-2\gamma} \; , \\
\delta &=& \sigma e^{-2\gamma} \; .
\end{eqnarray}
\end{subequations}
The field equations then become:
\begin{subequations} \label{skewdiv}
\begin{eqnarray}
&& (\eta r^{2})_{,r}\sin\theta + (\delta \sin\theta)_{,\theta} =0
 \; , \label{skewdiv1} \\
&& \delta_{,u}+\kappa_{,r}-(Ur^{2} \eta)_{,r}=0 \; , \label{skewdiv2} \\
&& r^{2}\eta_{,u} \sin\theta\,  + r^{2} (U\eta \sin\theta)_{,\theta}
-(\kappa \sin\theta)_{,\theta}=0 \; . \label{skewdiv3}
\end{eqnarray}
\end{subequations}
Since $g_{\mu\nu}$ must transform under the Poincar\'{e} group when
$r \rightarrow \infty$, we require that $\eta, \kappa$ and $\delta$ have
the following expansions in $1/r$.
\begin{subequations}
\begin{eqnarray}
\eta&=& n_{1}/r + n_{2}/r^{2} + ...\; , \\
\delta &=& d_{0}+d_{1}/r+d_{2}/r^{2}+...\; ,\\
\kappa &=& k_{0}+k_{1}/r +k_{2}/r^{2}+...\; .
\end{eqnarray}
\end{subequations}
We begin by solving for the coefficients of $h$ at lowest order in $1/r$.
It is only when we find these coefficients that we can be sure that the next
terms in the expansion of $h$ are smaller than the $h^{2}$ terms that we have
neglected.

Equation (\ref{skewdiv2}) yields $d_{0\, ,u}=0$, and since
$d_{0}=0$ when the system passes through its equillibrium position, $d_{0}=0$
always. Equation (\ref{skewdiv1}) then gives $n_{1}=0$. Continuing to next
order, we find from equation (\ref{skewdiv1}) that
$(d_{1} \sin\theta )_{,\theta}=0$. This equation for $d_{1}$ demands
$d_{1}=0$ in order to get a solution that is regular on the polar axis. The
next equation in the set, (\ref{skewdiv2}), yields $k_{1}=d_{2\, ,u}$ while
the equation (\ref{skewdiv3}) yields $n_{2\, ,u}=\sin^{-1}\theta
(k_{0}\sin\theta)_{,\theta}$.

To gain some understanding of the expansions thus far, it is instructive
to calculate the NGT charge, $L^{2}$, of the system
\begin{equation}
L^{2}\equiv {1 \over 4\pi}\int \left(\sqrt{-g}g^{[0\nu]}
\right)_{,\nu}d^{3}x={1 \over 2}\int_{0}^{\pi}n_{2}\sin\theta\, d\theta
=<n_{2}> \; ,
\end{equation}
where the brackets $<>$ denote the angular average.
To maintain a notation consistent with other work in NGT we define $n_{2}=
l^{2}Q(u,\theta)$ where $l^{2}$ is a constant (with respect to $r,\phi,u$
and $\theta$) with the dimensions of a $[{\rm length}]^{2}$ and is identified
as the usual NGT charge, and $<Q>=1$ for times when the system passes through
its equilibrium position. Notice that $L^{2}_{,u}=0$ by dint of the
condition
$n_{2\, ,u}=\sin^{-1}\theta (k_{0}\sin\theta)_{,\theta}$. This expresses the
important fact that the NGT charge of a body cannot be radiated away, and
is analogous to the situation in electromagnetism where the electric charge of
an antenna does not change.

We can now use the results of the lowest order expansion to find out how
far we can safely expand $h$ in inverse powers of $r$ before needing to
incorporate $h^{2}$ terms. This can easily be accomplished by studying the
exact form of $g^{\mu\nu}$ and $g$. We find that the linear order can
accurately determine $\eta, \kappa$ and $\delta$ as follows:
\begin{eqnarray}
\eta&=& Ql^{2}r^{-2}+Rl^{3}r^{-3}+Sl^{4}r^{-4} \; , \\
\kappa&=& Il+Jl^{2}r^{-1}+Kl^{3}r^{-2} \; , \\
\delta&=& Al^{3}r^{-2}+Bl^{4}r^{-3} \; ,
\end{eqnarray}
(Only even powers of $l$ will appear in physical quantities such
as the curvature tensor). The functions of integration are dimensionless
functions of both $u$ and $\theta$. They satisfy:
\begin{subequations}
\begin{eqnarray}
Q_{,u}&=&{(I\sin\theta)_{,\theta} \over \sin\theta} \; , \\
J&=&l A_{,u} \; , \label{aa} \\
R&=& {(A\sin\theta)_{,\theta} \over \sin\theta} \; , \label{bb} \\
R_{,u}&=& {(J\sin\theta)_{,\theta} \over l\sin\theta} \; , \label{cc}\\
K&=&{l B_{,u} \over 2} - {(c_{,\theta}+2c\cot\theta) \over l} \; ,\label{dd} \\
S&=& {(B\sin\theta)_{,\theta} \over 2\sin\theta} \; , \label{ee} \\
S_{,u}&=&{1 \over \sin\theta} \left({\sin\theta\left(c_{,\theta}
+2c\cot\theta+lK\right) \over l^{2} }\right)_{,\theta} \; . \label{ff}
\end{eqnarray}
\end{subequations}
Notice that (\ref{bb}), (\ref{cc}) can be combined to give
(\ref{aa}) and that (\ref{ee}), (\ref{ff}) can be combined to give
(\ref{dd}). This is a consequence of the identity (\ref{ident}).

Substituting the first terms in the above expansions into (\ref{srex}) gives:
\begin{subequations}
\begin{eqnarray}
W_{[2,0]}&=&-{3l \over 4r^{2}}(2lJ+2cI+lQ_{,\theta})_{,u} +O(r^{-3})\; , \\
W_{[0,1]}&=&O(r^{-4})\; ,\\
W_{[1,2]}&=&{3l \over 2r^{3}}(2lJ+2cI+lQ_{,\theta})+O(r^{-4}) \; .
\end{eqnarray}
\end{subequations}
The identity $W_{\{[\mu,\nu],\rho\}}=0$ then gives
\begin{eqnarray}
&&(2l^{2}J+2clI+l^{2}Q_{,\theta})_{,u}=0 \nonumber \\
&\Rightarrow & 2l^{2}J+2clI+l^{2}Q_{,\theta}=l^{2}f(\theta)\; .
\end{eqnarray}
The field strength $F_{\mu\nu\rho}$ is
\begin{equation}
F_{012}={l^{2}f(\theta) \over r^{2} } +O(r^{-3}) \; ,
\end{equation}
and since $F_{012}=0$ when the system passes through its equilibrium
position, $f(\theta)=0$.

Our results can be summarised as folows:
\begin{subequations}
\begin{eqnarray}
h^{[\mu\nu]}&=&O(r^{-2})+... \; , \label{fff} \\
W_{[\mu,\nu]}&=&O(r^{-3})+... \; , \label{fws} \\
F_{[\mu\nu\rho]}&=&O(r^{-3})+... \label{fs} \; .
\end{eqnarray}
\end{subequations}

\section{Calculating the Radiation Flux}

The rate of energy loss for an isolated body in NGT is given by
\begin{equation}
{dE \over dt} = -R^{2} \oint t^{0i}\hat{n}_{i}\, d\Omega \; ,
\end{equation}
where the integration is over a sphere of radius $R$ in the wave zone,
$\hat{n}_{i}$ is an outward pointing unit vector, and $t^{\mu\nu}$ is
the energy-momentum pseudo-tensor. The skew contribution to $t^{\mu\nu}$
is given by \cite{DDM}:
\begin{eqnarray}
t^{(\mu\nu)}_{{\rm skew}}&=& ({1\over 2}F^{\mu\alpha\beta}F^{\nu}_{\;\, \alpha
\beta}-{1 \over 12}\, ^{GR}g^{\mu\nu}F^{2}) \nonumber \\
&& +{2 \over 3}\left(2 h^{[\mu\alpha]}\, ^{GR}g^{\nu\beta}W_{[\beta,\alpha]}
+ 2 h^{[\nu\alpha]}\, ^{GR}g^{\mu\beta}W_{[\beta,\alpha]}
-\, ^{GR}g^{\mu\nu}h^{[\alpha\beta]}W_{[\alpha,\beta]}\right) \; .
\end{eqnarray}
Inserting our expressions (\ref{fs}), (\ref{fff}) and (\ref{fws})
for $F_{\mu\nu\alpha}, h^{[\mu\nu]}$ and
$W_{[\mu,\nu]}$, we find that $t^{(\mu\nu)}_{{\rm skew}}=O(r^{-5})$ so
there is no skew contribution to the radiation flux in NGT. The only
non-vanishing gravitational radiation flux in NGT comes from the
leading order terms of the symmetric sector, which reproduce the usual
GR radiation formula.

As explained in the introduction, the linearized wave equations allow
us to obtain the radiation pattern of {\em any generic source} by linearly
superimposing a suitable combination of differently aligned axi-symmetric
solutions. Since we know that the radiation flux is identical in NGT
and GR for the axi-symmetric case, we know that the flux will be identical in
general. This line of reasoning can be rigorously supported by generalizing
the work of Sachs \cite{Sachs}, who found that the axi-symmetric wave
solution of Bondi, van der Burg and Metzner \cite{BBM} contained all
the essential features needed to describe gravitational radiation from
a generic source of no particular symmetry.

\section{Conclusions}
We have proved that the flux of gravitational radiation from an isolated
source in NGT is identical to that found in GR, a result that clearly
refutes the claims made by Damour, Deser and McCarthy.

In many respects our result is not unexpected. The two conserved sources
for an isolated system in NGT are $L^{2}$, the NGT charge, and $T^{\mu\nu}$,
the energy momentum tensor. Since $L^{2}$ has the dimensions of a $[{\rm
length}]^{2}$, while the constant $M$ in the GR sector has the
dimensions of a length, we immediately expect the skew sector to be shorter
ranged than the symmetric sector. Moreover, from the viewpoint of the
phenomenological model where $L^{2}$ is taken to be proportional to the
fermion number, we expect that $L$ will be a constant since the fermion number
of an isolated body is a constant. This was borne out in our solution.
Thus, there will be no dimensionless quantity $L^{2}_{,u}$ available
in the skew sector unlike the symmetric sector
where the dimensionless quantity $M_{,u}$ is available. Using dimensional
analysis, we can then predict the form of the Riemann tensor, which must
have the dimensions of $[{\rm length}]^{-2}$ and contains at most two
time derivatives. The symmetric sector can contribute super-radiative
$1/r$ terms and radiative $1/r^{2}$ terms, while the skew sector can
only contribute non-radiative $1/r^{4}$ (and lower) terms.

\vspace{.5in}
{\bf Acknowledgements}

This work was supported by the Natural Sciences and Engineering
Research Council of Canada. One of the authors (NJC) is
grateful for the support provided by a Canadian Commonwealth
Scholarship.

\end{document}